# Design Optimization of a Small-animal SPECT System Using LGSO Continuous Crystal and a Micro Parallel-hole Collimator

[1]Joong Hyun Kim, [2]Mikiko Ito, [2]Soo Mee Kim, [3]Seong Jong Hong, [2,4]Jae Sung Lee,

[1]Center for Ionizing Radiation, Division of Metrology for Quality of Life, Korea Research Institute of Standards and Science

[2]Department of Nuclear Medicine, Seoul National University College of Medicine

[3]Department of Radiological Science, Eulji University

[4]Department of Biomedical Sciences and Institute of Radiation Medicine, Medical Research Center Seoul National University College of Medicine

Name and address of correspondence and reprint request:

**Jae Sung Lee, PhD**

Department of Nuclear Medicine, Seoul National University College of Medicine

101 Daehak-Ro, Jongno-Gu, Seoul 110-744, Korea

Phone: +82-2-2072-2938, Fax: +82-2-745-2938, e-mail: jaes@snu.ac.kr

Total number of words in the manuscript: 4,748

**Short Running Headline**: Design optimization of LGSO SPECT

**Abstract**

**Purpose:** The aim of this study was to optimize the design of a monolithic LGSO scintillation crystal and micro parallel-hole collimator for the development of a small-animal single photon emission computed tomography (SPECT) system with compact size, low-cost and reasonable performance through Monte Carlo simulation.

**Methods:** $L_{0.9}$GSO crystals with surface area of 50 mm × 50 mm were investigated for the design optimization. The intrinsic detection efficiency, intrinsic spatial resolution, and intrinsic energy resolution of crystals were estimated for different crystal thicknesses and photon energies (using I-125 and Tc-99m sources). Two kinds of surface treatments (providing polished and rough surfaces) were compared by optical photon simulation. The standard deviation of the angle between a micro-facet and the mean surface was set to 0.1 and 6.0 for polished and rough surfaces, respectively. For comparison, the intrinsic performance of NaI(Tl) was also investigated. A multi-photomultiplier tube was designed with 16 × 16 anode pixels having size of 2.8 mm × 2.8 mm and pitch of 3.04 mm, and a 1.5 mm thickness glass window. The length of the micro collimator was also optimized. Finally, the performance of the SPECT system was assessed and an ultra-micro hot spot phantom image was obtained in simulation.

**Results:** The 1-mm-thick LGSO was sufficient to detect most incident photons from I-125 but a thickness of 3 mm was required for Tc-99m imaging. Polished crystal yielded better intrinsic spatial resolution (~540 μm) and lower light output than rough crystal. Energy resolutions of I-125 and Tc-99m were ~36.9% and ~19.1%. With the optimized collimator length, spatial resolution of ~1 mm and sensitivity of ~100 cps/MBq were achieved with a four-head SPECT system. A hot rod with a diameter of 1.0 mm was resolved in the SPECT image of ultra-micro hot spot phantom.

**Conclusion:** A small-animal SPECT system having compact size and low cost was designed. Using a thin monolithic crystal and micro collimator, high spatial resolution and high sensitivity were achieved.

**INTRODUCTION**

Nuclear medicine imaging techniques provide the spatiotemporal biodistribution of radiolabeled molecular imaging probes related to various diagnostic information for a wide range of disease statuses. In addition, pre-clinical molecular imaging methods based on nuclear medicine technologies play important roles in various areas, such as the development of new drugs and radiopharmaceuticals. Small-animal dedicated single photon emission computed tomography (SPECT) systems with fine spatial resolution and high sensitivity are now widely used and regarded as one of the most important types of *in vivo* molecular imaging tools. In particular, small-animal SPECT systems allow the acquisition of rodent images with finer spatial resolution than small-animal positron emission tomography systems that have intrinsically limited spatial resolution originating from the positron range and non-colinear photon annihilation.

Iodine-125 (I-125) is a widely used radioisotope in nuclear medicine. However, its application is limited to *in vitro* radioimmunoassays, anti-body studies and autoradiography because I-125 emits low-energy photons (less than 35.5 keV X-rays and gamma rays with energy less than 35.5 keV). Accordingly, clinical SPECT systems with a high signal trigger threshold usually cannot use I-125.

Currently available commercial pre-clinical SPECT systems mainly use pinhole collimators to achieve a high magnification factor and fine spatial resolution (as good as ~0.35 mm). However, the high magnification factor requires large-area scintillation cameras and close proximity between the imaging object and pinhole collimator, resulting in a bulky and expensive SPECT system but a limited scan field of view. Low sensitivity is another disadvantage of the pinhole collimator but has been overcome using multiple detector heads and multiple pinholes [1,2].

In this study, we designed a small-animal SPECT system with compact size, sufficiently large field of view, low cost and reasonable sensitivity and spatial resolution. A parallel-hole collimator with micro architecture was adopted for this system. Because the magnification factor of the parallel-hole collimator is 1, a large-area scintillation camera is not required to obtain a sufficient field of view. Reasonable system performance can be achieved by the micro architecture of the parallel-hole collimator. In this study, we performed a Monte Carlo simulation using the Geant4 application for tomographic emission (GATE) simulation code to optimize the design of the proposed SPECT system [3].

**MATERIALS AND METHODS**

**Optimization of Intrinsic Performance**

The intrinsic performance of a gamma camera is mainly determined by the scintillation crystal and photomultiplier tube (PMT). For the readout of scintillation light, we selected an H9500 PMT (Hamamatsu Photonics K.K., Japan), which is a position-sensitive multi-anode flat panel PMT that has the advantages of compact size, large effective area, and small pixel size. The effective area of the H9500 is 49 mm × 49 mm. A total of 256 anodes are arranged in a 16 × 16 array. The size of each anode is 2.8 mm × 2.8 mm, and the pitch between neighboring anodes is 3.04 mm [4]. Anodes have thickness of 0.1 mm, and are covered by glass of 1.5 mm thickness.

An $L_{0.9}$GSO scintillation crystal was selected to optimize the SPECT performance. The LGSO has high density of ~7.4 g/cm$^3$ and is a non-hydroscopic crystal. However, LGSO has lower light output than NaI(Tl) and yields intrinsic radioactivity from Lu-176. The intrinsic performance of LGSO-based detectors is compared with the intrinsic performances of NaI(Tl)-based detectors in this study. The major properties of these crystals are listed in Table 1.

To optimize the intrinsic performance of the crystal, the crystal thickness and surface treatments were investigated. Crystal thickness is a dominant factor determining the intrinsic detection efficiency and spatial resolution of the scintillation camera. There is a trade-off between efficiency and resolution. A thicker crystal captures more incident radiation but provides worse spatial resolution owing to the wider spread of the optical photons in the crystal.

The crystal surface treatment is another factor that determines intrinsic performance.

A thin monolithic crystal has six surfaces to be treated: two main surfaces with large area and four narrow surfaces at the sides. A cutting surface of the scintillation crystal that has not been mechanically or chemically polished is referred to as "rough as sliced (or rough)". On the rough surface, the standard deviation of the angle between the mean surface and the micro facet of the crystal is higher than that for a polished crystal surface.

We designed the GATE simulation environment as follows. A thin monolithic crystal of 1 or 3 mm thickness is coupled with the H9500 PMT, and a thin optical grease layer of 0.1 mm is applied between the crystal and PMT. The other side of the crystal is covered by an optical reflector. The entrance window of the PMT is a monolithic glass layer of 1.5 mm thickness. Under the glass layer, there are 256 photocathode pixels that have a total area of 2.8 mm × 2.8 mm. The packing fraction of these anode pixels is 89%. The ~11% of optical photons that do not reach the photocathode pixels are absorbed or reflected. In a real situation, approximately 89% of optical photons have a chance of being converted into electrons at photocathode pixels, and these electrons are dispersed in a vacuum layer of 2.5 mm thickness. At the end of the vacuum layer, these electrons reach the first dynodes. However, the photoelectric effect cannot be simulated easily in GATE Monte Carlo code. For the convenience of simulation, we adopt a second glass layer rather than a vacuum layer and assume that there is no photoelectric effect between the first and second layers. The optical photons disperse again in the second glass layer, and this dispersion is assumed as being similar to the photoelectric effect. The optical photons passing through the second glass layer are absorbed at the end of the second layer, and the number of absorbed photons is recorded for each pixel.

To evaluate the performance of intrinsic properties, we assessed the intrinsic efficiency, spatial resolution and energy resolution, optical photon light output, and displacement error of the radiation interaction position. In all performance evaluations, ideal small-beam sources were used because these evaluations focused on intrinsic performance. Radiation sources were 140.5 keV in the Tc-99m simulation and 35.5 keV in the I-125 simulation. The highest gamma-ray energy of I-125 is 35.5 keV. The number of absorbed optical photons was recorded for each pixel, and the position of incident radiation was calculated according to a center-of-mass calculation. Evaluations were performed for rough and polished surfaces. The standard deviation of the angle between the mean surface and a micro facet (sigma alpha) was 0.1 for the polished surface and 6.0 for the rough surface [6].

The intrinsic efficiency was assessed for two crystal thicknesses and two crystal surface treatments of Tc-99m and I-125. The center of the crystal was irradiated with $10^6$ photons. The percentages of valid events in the energy window were calculated. The energy windows were 40% for I-125 and 20% for Tc-99m and differed because the intrinsic energy resolutions were different for the two types of crystal. The intrinsic efficiencies were reported for each condition.

The crystal was irradiated with $10^4$ radiations from the center to the corner of the crystal. The positions of detected radiations were recorded as a 128 × 128 matrix, with the pixel size being 380 μm × 380 μm. For one quadrant, radiations were irradiated from [0.19, 0.19] mm to [24.13, 24.13] mm in steps of 2.66 mm (i.e., [0.19 + 2.66$n$, 0.19 + 2.66$n$] for $n$ = 0, 1, …, 9).

The intrinsic spatial resolution was measured by Gaussian fitting of the spatial distribution of the incident radiation and was reported as the full width at half maximum (FWHM) in units of millimeters. The intrinsic energy resolution was assessed as the FWHM of the distribution of the optical light output. And ratio of the FWHM and optical light output at peak is the intrinsic energy resolution.

Usually, the position of radiation interaction in monolithic crystal at the edge or corner tends to move to the center. This error is referred to as displacement error. The displacement error was calculated as the distance between the actual irradiated position and calculated interaction position in image.

**Optimization of the Collimator Performance**

The collimator performance is another decisive factor of the overall performance of the SPECT system. A parallel-hole collimator having micro architecture was designed. The area of the collimator is 52 mm × 52 mm so as to cover the area of the scintillation crystal. The collimator consists of 128 × 128 square holes arranged in a square matrix. The area of each hole is 340 μm × 340 μm and the pitch between holes is 380 μm. The septal thickness is 40 μm. The collimator material is tungsten.

The collimator performance was optimized for the collimator length. There is a trade-off between the spatial resolution and sensitivity of a collimator. A longer collimator yields higher spatial resolution but lower sensitivity. The collimator length was optimized to yield reasonable resolution and sensitivity with consideration of the intrinsic performance of the crystal.

To optimize the collimator length, the projection image of a point source at a distance of 26 mm from the collimator surface was acquired with various collimator lengths ranging from 10 to 20 mm. At the end of the collimator, we placed an ideal radiation absorber with 100% absorption efficiency and spatial resolution of 0 mm because the simulation focused on the collimator performance. The point source generated $10^7$ radiations and a planar projection image was acquired. The collimator sensitivity was calculated from the total number of recorded counts in the projection image and the number of generated radiations. The spatial resolution of the collimator was acquired by Gaussian fitting of the line profile through the point source in the projection image.

**Evaluation of Intrinsic Activity**

LGSO crystal has intrinsic radioactivity due to the presence of lutetium-176. The natural abundance of Lu-176 is 2.59% of all lutetium and the half-life of Lu-176 is 3.78 $\times\ 10^{10}$ years. Calculated Lu-176 activities of crystals 1 and 3 mm thick are ~655 and ~1965 Bq, respectively. The intrinsic radioactivity per square centimeter was measured within two energy windows, 40% for 35.5 keV (I-125) and 20% for 140.5 keV (Tc-99m). The intrinsic radioactivity was compared with a flood source image for radioactivity of 2.5 MBq/cm$^2$ (67.6 μCi/cm$^2$), and the background activity level was assessed. To obtain the flood source image, the optimized collimator and crystal described above were used.

**SPECT Performance**

To estimate the performance of the entire SPECT system, we designed a compact four-head SPECT system using the above optimized collimator and crystal. A field of view of approximately 5 cm in transaxial and axial directions can be obtained using the four-head system. The sensitivity and planar spatial resolution of the system were assessed according to the intrinsic crystal and collimator performances. To evaluate the spatial resolution of the system, an ultra-micro hot spot phantom (model: ECT/HOT/UMMP, Data Spectrum Corp. NC, USA) was designed in GATE simulation. The ultra-micro phantom has a diameter of 2.8 cm and height of 2.8 cm and consists of six hollow channels having diameters of 0.75, 1.0, 1.35, 1.7, 2.0, and 2.4 mm.

A 100-MBq (2.7-mCi) quantity of I-125 was distributed in the hollow channels of the ultra-micro phantom. The energy spectrum of I-125 was simplified to four photon energies from 27.3 to 35.5 keV with adequate yields. One-hundred and twenty projections were acquired in only 2 minutes (4 s for each projection) and the rotation angle was 3 degrees. Projections were reconstructed by filtered back-projection with a Hanning filter and ordered-subset expectation maximization with four subsets and four iterations.

**RESULTS**

**Intrinsic Performance**

  The energy spectra of LGSO under the conditions of a polished surface treatment and a crystal thickness of 3 mm are shown in Figure 1. The energy resolutions of I-125 and Tc-99m were approximately 36% and 18%. Therefore, the intrinsic efficiencies for each isotope were estimated in different energy windows, 40% for I-125 and 20% for Tc-99m. The intrinsic efficiency was mainly determined by the crystal type and thickness, and not by the crystal surface treatment. The NaI(Tl) crystal had unacceptable efficiency for the gamma energy of Tc-99m because of its low density. The LGSO crystal, however, had high efficiency. A thickness of LGSO crystal of only 1 mm yielded perfect efficiency in the entire energy window for I-125. The detection efficiency of I-125 in the 40% energy window for the 1-mm-thick crystal was not notably different from that for the 3-mm-thick crystal. In the case of Tc-99m, the LGSO crystal of 1 mm thickness had low detection efficiency (<40%) whereas the LGSO of 3 mm thickness had detection efficiency of approximately 70% in the 20% energy window. Only the LGSO crystal of 3 mm thickness was suitable for both I-125 and Tc-99m SPECT imaging.

  The energy resolution for Tc-99m was approximately 18%, and there was no notable difference according to the surface treatment. However, the energy resolution worsened as the source position moved to the edge of the crystal. I-125 had energy resolution (approximately 36%) twice that of Tc-99m. The trends of the energy resolution of I-125 were similar to those of Tc-99m in terms of the change in energy resolution with changes in surface treatment and source position (Figure 2).

The optical light output depended on the surface treatment and source position (Figure 3). A rough surface yielded a higher light output than a polished surface (by approximately 3%–14% for 3-mm-thick crystal). The optical light output decreased as the source position moved to the edge. For the rough surface, the optical light output at the edge was approximately 20%–28% lower than that at the center.

The intrinsic spatial resolution of the 3-mm-thick rough crystal was 0.57–1.04 and 1.03–1.51 mm for Tc-99m and I-125, respectively (Figure 4). When the sources were positioned at the corner of the crystals, the intrinsic spatial resolution of the 3-mm-thick rough crystal was approximately 5.5% and 9.9% better than that of the 3-mm-thick polished crystal for Tc-99m and I-125 respectively.

The displacement error depended on the surface treatment. Two-dimensional flood images are shown in Figure 5. The polished crystal performed better in distinguishing point sources. The displacement error of the polished crystal was approximately 0.5 mm less than that of the rough crystal for both of Tc-99m and I-125.

**Collimator Performance**

Figures 6A and 6B show the sensitivity and spatial resolution of the collimator versus collimator length. The collimator sensitivity decreased from 45.3 to 11.6 cps/MBq as the collimator length increased from 10 to 20 mm. The collimator spatial resolution improved from 0.8 to 0.45 mm as the collimator length increased from 10 to 20 mm.

**Intrinsic activity**

The intrinsic activity induced by Lu-176 was assessed. For 3-mm-thick crystal, the background activities were 2.33 and 2.75 cps/cm$^2$ for the Tc-99m and I-125 energy windows, respectively. With the flood source of 2.5 MBq/cm$^2$, the detected activities were 79.9 and 107.1 cps/cm$^2$ for Tc-99m and I-125, respectively. The ratios of background activity over the total count were 2.9% and 2.6% for Tc-99m and I-125, respectively.

**SPECT performance**

The performances of the entire SPECT system can be estimated from the intrinsic and collimator performances. The planar spatial resolution of the system was expected to be 1.18 mm for I-125 and 1.05 mm for Tc-99m. The system sensitivity with four heads was expected to be 133 cps/MBq for I-125 and 86 cps/MBq for Tc-99m.

The micro phantom image reconstructed using ordered-subset expectation maximization with four subsets and four iterations is shown in Figure 7. Hot spot inserts as small as 1 mm could be resolved.

**DISCUSSION**

  Commercial micro SPECT systems that are currently available mainly use multi-pinhole collimators to achieve high spatial resolution and high sensitivity. With the pinhole collimator, high spatial resolution can be achieved using a large-area scintillation crystal and high magnification factor. However, the gamma camera based on a single pinhole collimator has unfavorable properties of low sensitivity, a small field of view and resolution degradation at an axially off-center position. The multi-pinhole collimator was proposed to overcome these disadvantages of the single-pinhole collimator.

  In this study, a micro SPECT system using a parallel-hole collimator with micro architecture and thin monolithic crystal was designed. This system provides a sufficiently large field of view for rodent imaging with a much smaller detector than that used in pinhole SPECT systems. The small detector has many advantages. A bench-top micro SPECT system can be developed with a volume of less than 1 $m^3$ and at reasonable cost. Because only a small amount of scintillation crystal material is required, expensive but high-quality crystals can be adopted.

  High sensitivity can be achieved with the parallel-hole collimator. However, the spatial resolution is worse than that of a pinhole collimator. It is possible to recover the resolution by iterative reconstruction using a point spread function or collimator−detector response.

  The crystal surface treatment affects the intrinsic properties of the detector. The incident radiation is converted into hundreds to thousands of optical photons in the crystal, and these optical photons travel through the crystal and reach the crystal surface. At the crystal surface, optical photons are reflected or absorbed or penetrate the surface. Two different surface treatments can be applied to the two large surfaces of crystals to obtain rough and polished surfaces. There is no notable difference in the absorption, penetration or reflection for the two surface treatments; however, the uncertainty in the reflection angle differs. We assigned different uncertainty levels for the surfaces (sigma alpha values were 0.1 and 6.0 for polished and rough surfaces, respectively). At the crystal surface, optical photons are reflected if the incidence angle is smaller than the critical angle determined by the reflective indexes of crystal and grease. On the polished surface, the incidence and reflection angles are similar, and therefore, the optical photon cannot escape through the large surface if the incidence angle is smaller than the critical angle. Only optical photons that reach the surface with an angle larger than the critical angle can escape and be detected by the PMT. In the case of the rough surface, however, the reflection angle is much dispersed after reflection. Therefore, the escape probability for the rough crystal is higher than that for the polished crystal. These properties were reflected in the experiment results of light output. The light output of the rough crystal was higher than that of the polished crystal. The energy resolution is related to the optical photon output. A high output yields better energy resolution, and the rough crystal therefore yielded better energy resolution than the polished crystal. The intrinsic spatial resolution can also be explained by the sigma alpha value. For the polished crystal, only the optical photons having an angle larger than the critical angle can escape the crystal, and they are detected in the proximal region not far from where the incident radiation is detected. In rough crystal, however, the escape sites of optical photons are widely dispersed. This explains why the intrinsic spatial resolution of the polished crystal was better than that of the rough crystal.

In this study, we finally selected the 3-mm-thick crystal over the 1-mm-thick crystal. For I-125 imaging, the 1-mm-thick crystal was sufficient. However, the detection of radiation from Tc-99m is more important in SPECT research because the Tc-99m is the most widely used isotope in SPECT. Although the detection rate of Tc-99m was only 70%, it was measured with a 20% energy window. For a crystal thicker than 3 mm, there is no increase in the detection rate but there is degradation of the spatial resolution. A thickness of 3 mm was thus optimal for the SPECT system.

The collimator length is another important factor that determines the spatial resolution and sensitivity of the system. A long collimator yields high spatial resolution but low sensitivity. In this study, we selected a collimator length of 15 mm that provided spatial resolution of ~1 mm and system sensitivity of ~100 cps/MBq.

**CONCLUSION**

We designed a small-animal SPECT system using a parallel-hole collimator and thin monolithic crystal. A compact and economic system can be produced on the basis of this design. The SPECT system had spatial resolution of ~1 mm and sensitivity of ~100 cps/MBq for Tc-99m and I-125 imaging, which is sufficient for the imaging of rodents. The system performance was verified in an experiment using an ultra-micro hot spot insert phantom.

**ACKNOWLEDGEMENTS**

This work was supported by grants from Seoul National University Hospital (04-2008-0280) and the National Research Foundation of Korea (NRF) funded by the Korean Ministry of Science, ICT and Future Planning (NRF-2014M3C7034000, NRF-2013R1A1A2013175).

**TABLE**

Table1. Major properties of NaI(Tl) and $L_{0.9}$GSO crystal are listed. The LGSO crystal have higher density and shorter decay time than NaI(Tl) crystal and is not hydroscopic. However, photon yield of LGSO is lower than NaI(Tl).

| Property | NaI(Tl) | $L_{0.9}$GSO |
|---|---|---|
| Density ($g/cm^3$) | 3.67 | 7.3 |
| Decay time (nsec) | 230 | 41 |
| Photon yield (per keV) | 38 | 23 |
| Hydroscopic | Yes | No |
| Peak emission (nm) | 410 | 420 |
| Intrinsic activity | No | Yes (Lu-176) |

Table 2. Intrinsic performanc of LGSO crystal for eight different conditions. In the case of the rough crystal, intrinsic energy resolution and efficiency were slightly better than polished crystal, however, spatial resolution was notably inferior. One-millimeter-thick crystal was inadequate for 140.5 keV photon because of low detection efficiency. Crystal in condition of polished and 3-mm-thick was selected for the SPECT system.

| Radiation Energy | Surface Treatment | Crystal Thickness | Intrinsic Energy Resolution (%) | Intrinsic Efficiency (40%, 20%)* (cps/MBq) | Intrinsic Spatial Resolution (mm) |
|---|---|---|---|---|---|
| 35.5 keV I-125 | Rough | 1 mm | 35.9 | 79.1 | 0.90 |
| | | 3 mm | 35.9 | 80.9 | 1.07 |
| | Polished | 1 mm | 36.3 | 78.0 | 0.55 |
| | | 3 mm | 36.9 | 77.6 | 0.74 |
| 140.5 keV Tc-99m | Rough | 1 mm | 18.4 | 39.2 | 0.50 |
| | | 3 mm | 18.6 | 70.4 | 0.61 |
| | Polished | 1 mm | 18.8 | 38.9 | 0.36 |
| | | 3 mm | 19.1 | 67.6 | 0.54 |

* Energy windows were 40% for 35.5 keV and 20% for 140.5 keV.

**FIGURE LEGENDS**

Figure 1. Energy spectra of 3-mm-thick LGSO crystal. (A) polished surface, 35.5 keV photon, (B) rough surface, 35.5 keV photon, (C) polished surface, 140.5 keV photon, (D) rough surface, 140.5 keV. The energy resolutions of 35.5 keV and 140.5 keV were about 36% and 18%, respectively. The light output of rough surface was higher than polished one.

Figure 2. Energy resolution of 3-mm-thick LGSO crystal. There is no notable difference between polished and rough surface. Energy resolution of 140.5 keV is better than that of 35.5 keV because of light output difference.

Figure 3. Optical light output of 3-mm-thick LGSO crystal. The optical light output of 140.5 is about 4 times higher than that of 35.5 keV because of energy difference. For the rough surface, the optical light output is slightly higher than polished one.

Figure 4. Intrinsic spatial resolution of 3-mm-thick LGSO crystal. At center, the intrinsic spatial resolution was ~0.55 mm and ~1 mm for 140.5 keV and 35.5 keV photons, respectively. The intrinsic spatial resolution was worse at the edge position than center position.

Figure 5. The two dimensional flood images of 3-mm-thick LGSO crystal. (A) polished surface, 35.5 keV photon, (B) rough surface, 35.5 keV photon, (C) polished surface, 140.5 keV photon, (D) rough surface, 140.5 keV. The displacement error of polished

crystal at the edge position was smaller than rough crystal.

Figure 6. The results of collimator performance according to collimator length. (A) The collimator sensitivity (cps/MBq), (B) Collimator spatial resolution.

Figure 7. The SPECT image of the ultra-micro hot spot phantom. One hundred twenty planar projection images were reconstructed by ordered-subset expectation maximization with four subsets and four iterations. Hot spot inserts as small as 1 mm could be resolved.

Figure 1

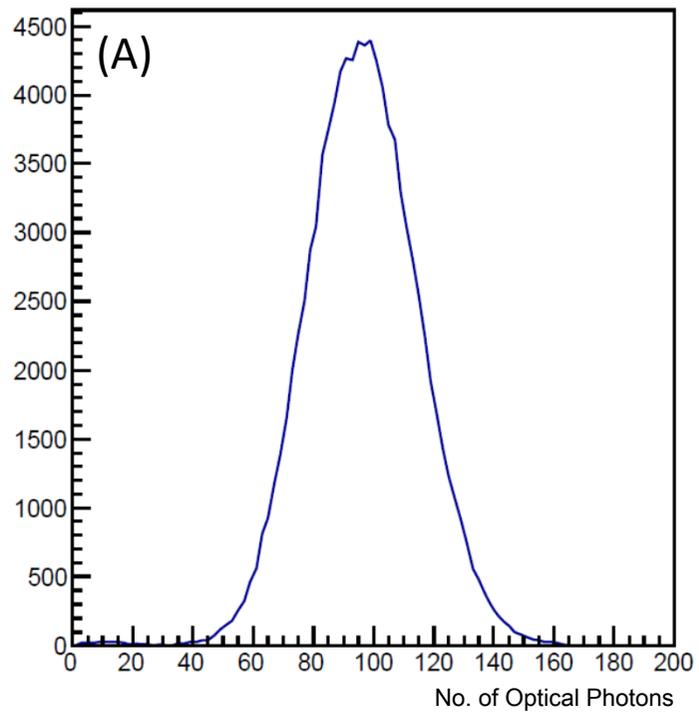
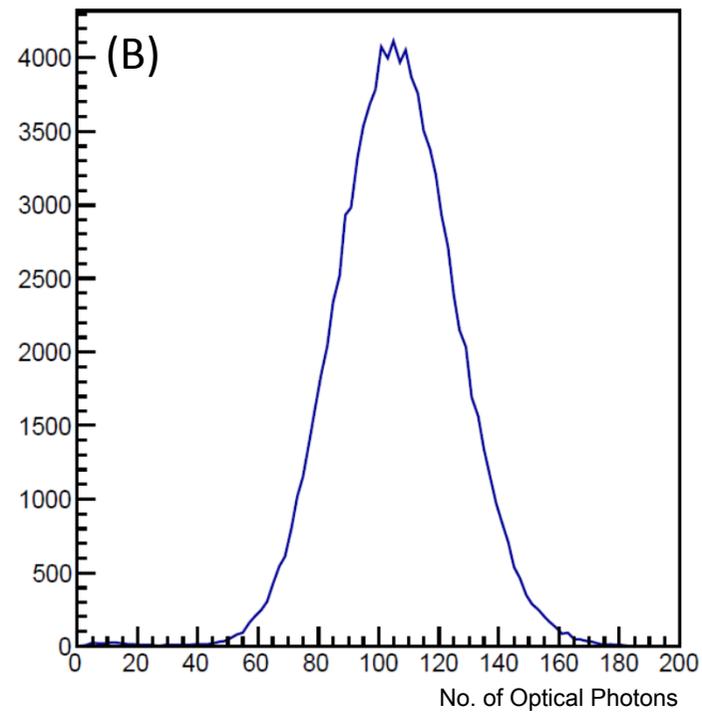
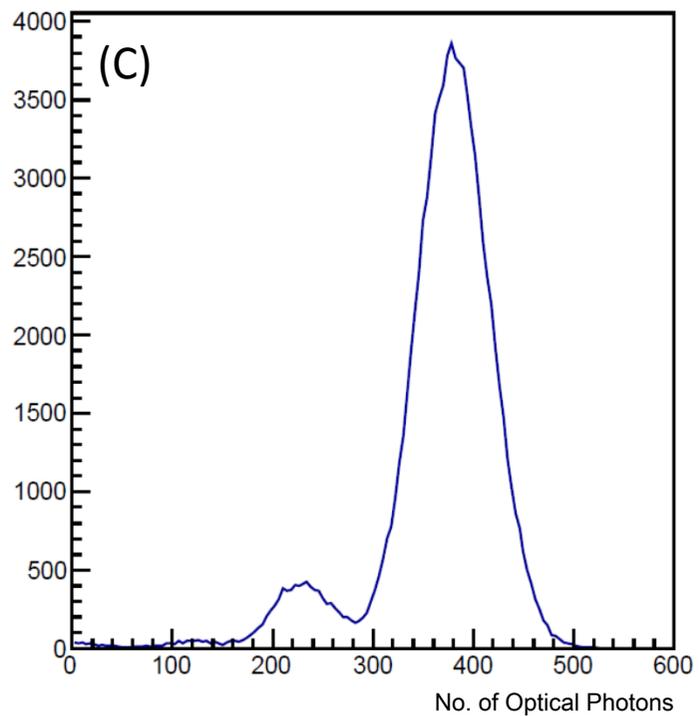
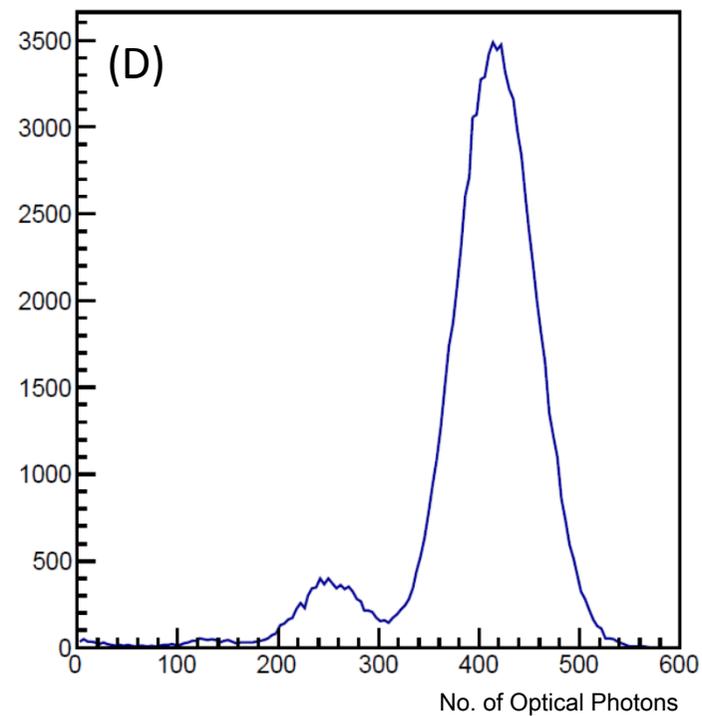

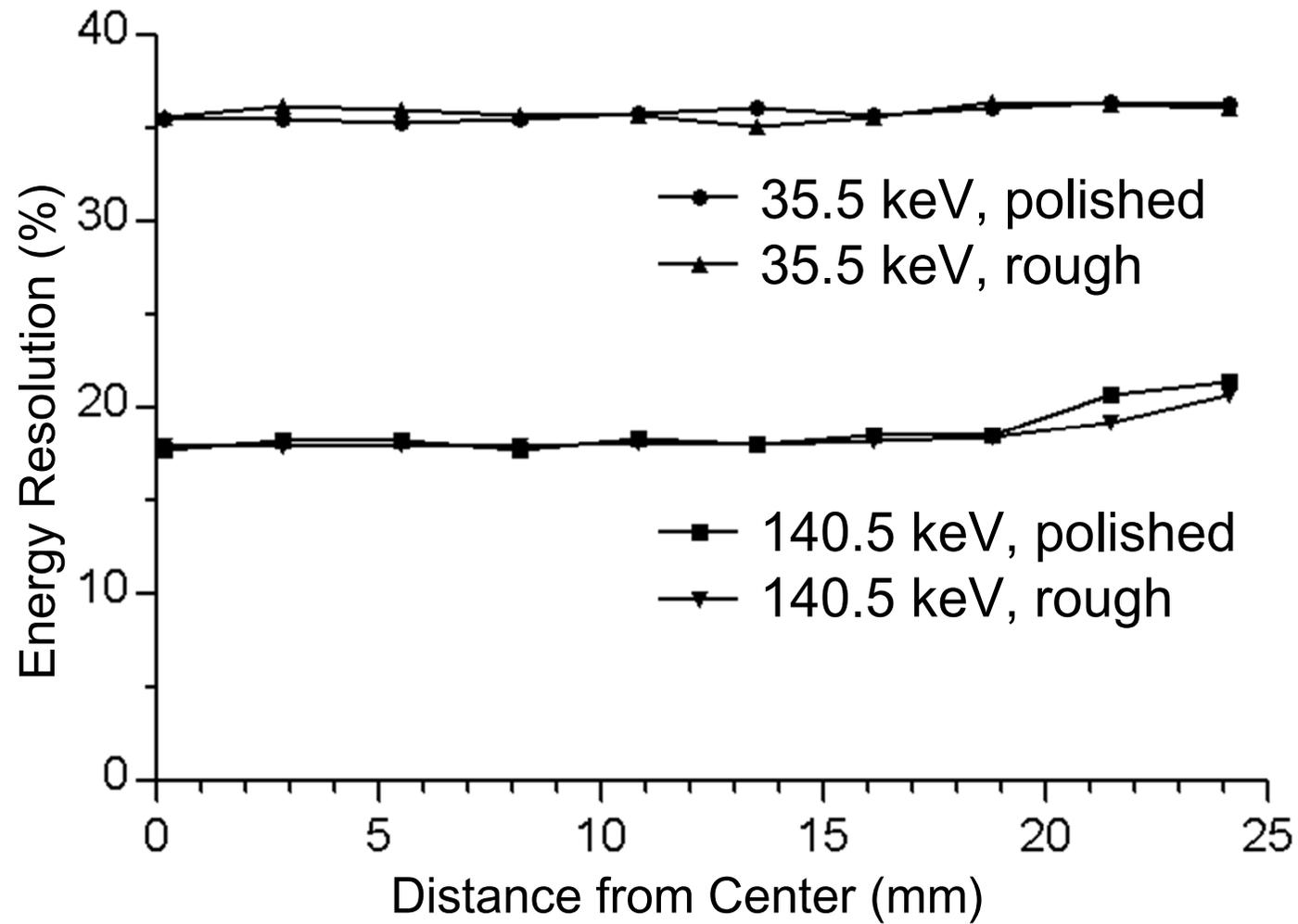

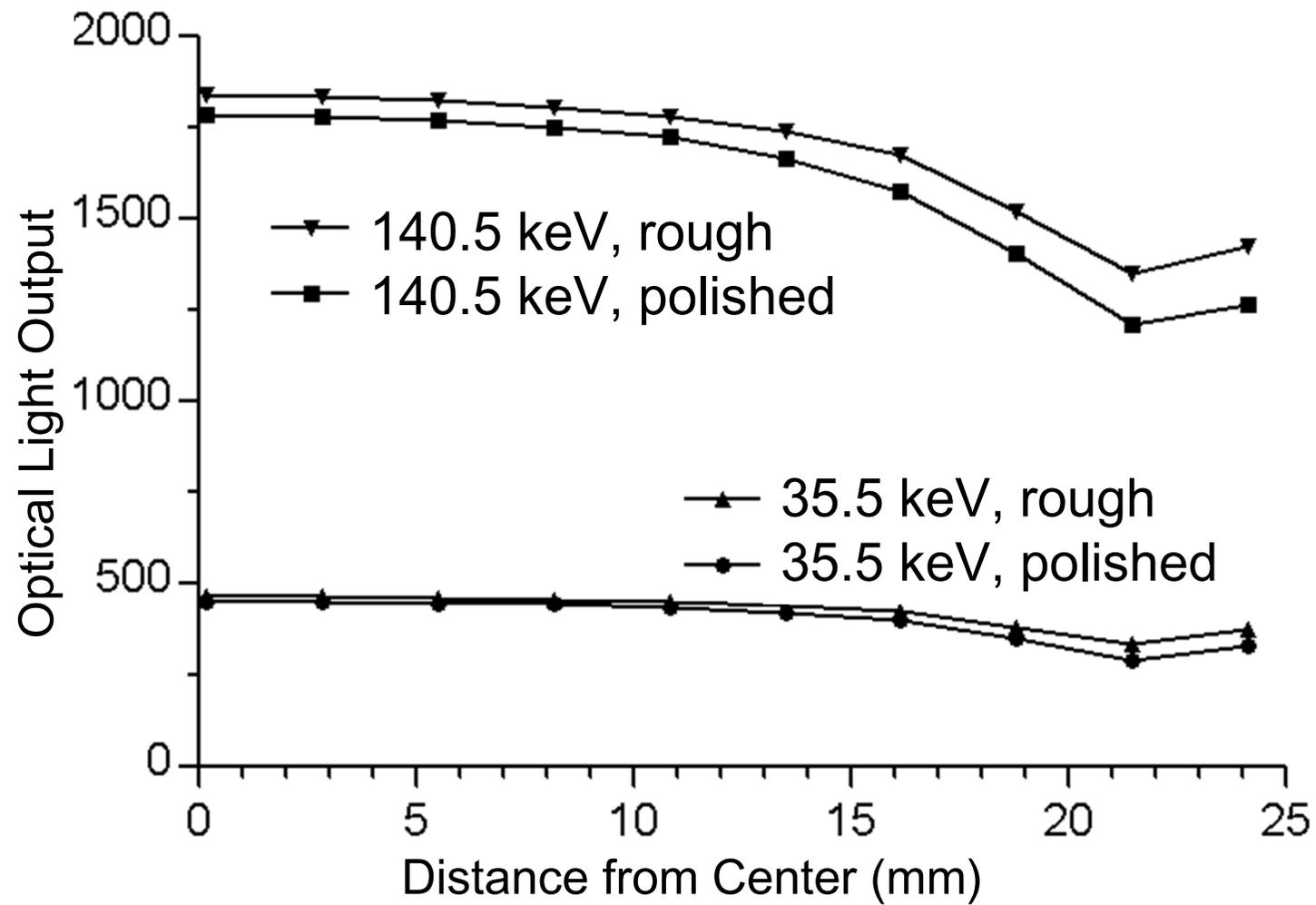

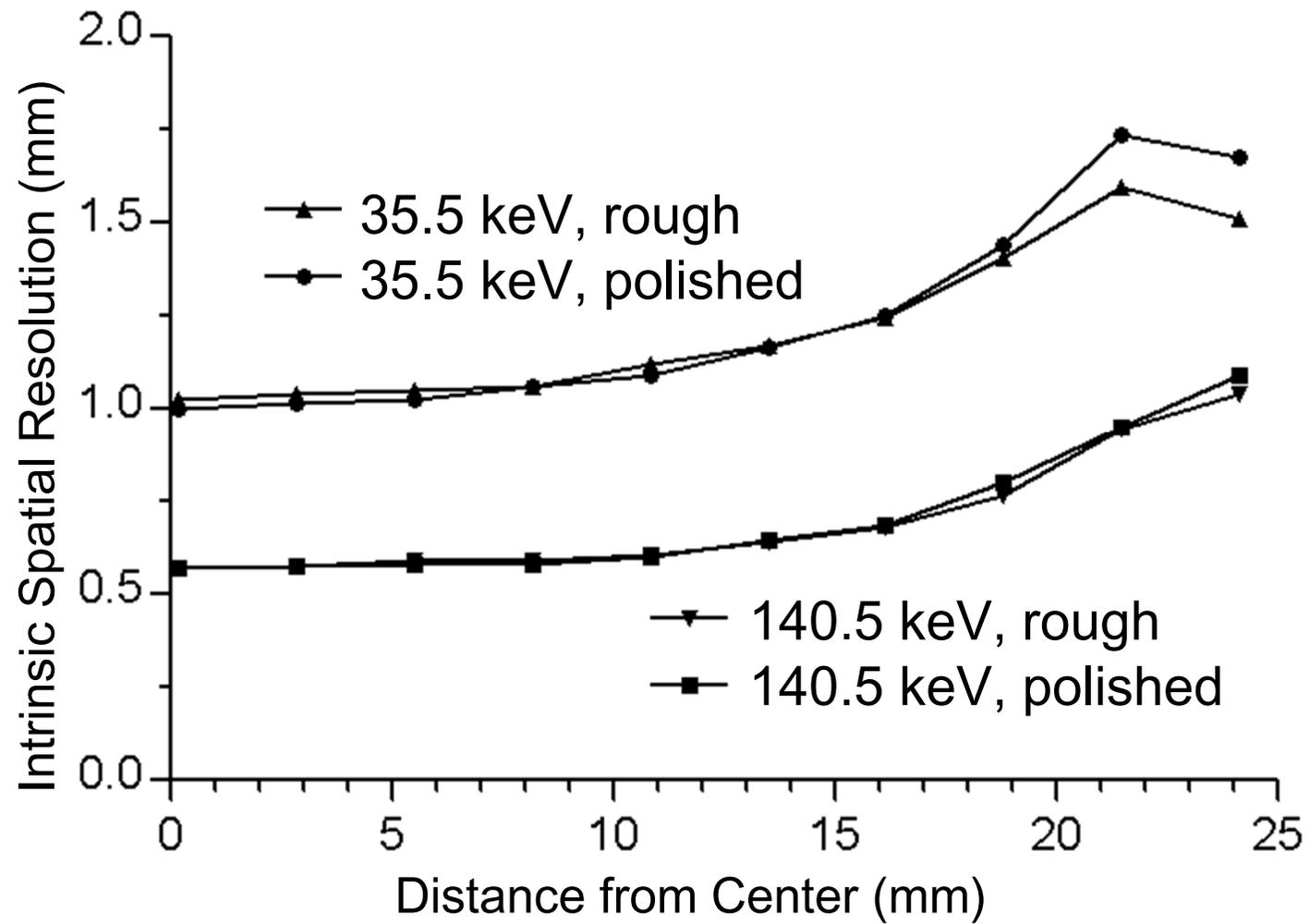

Figure 5

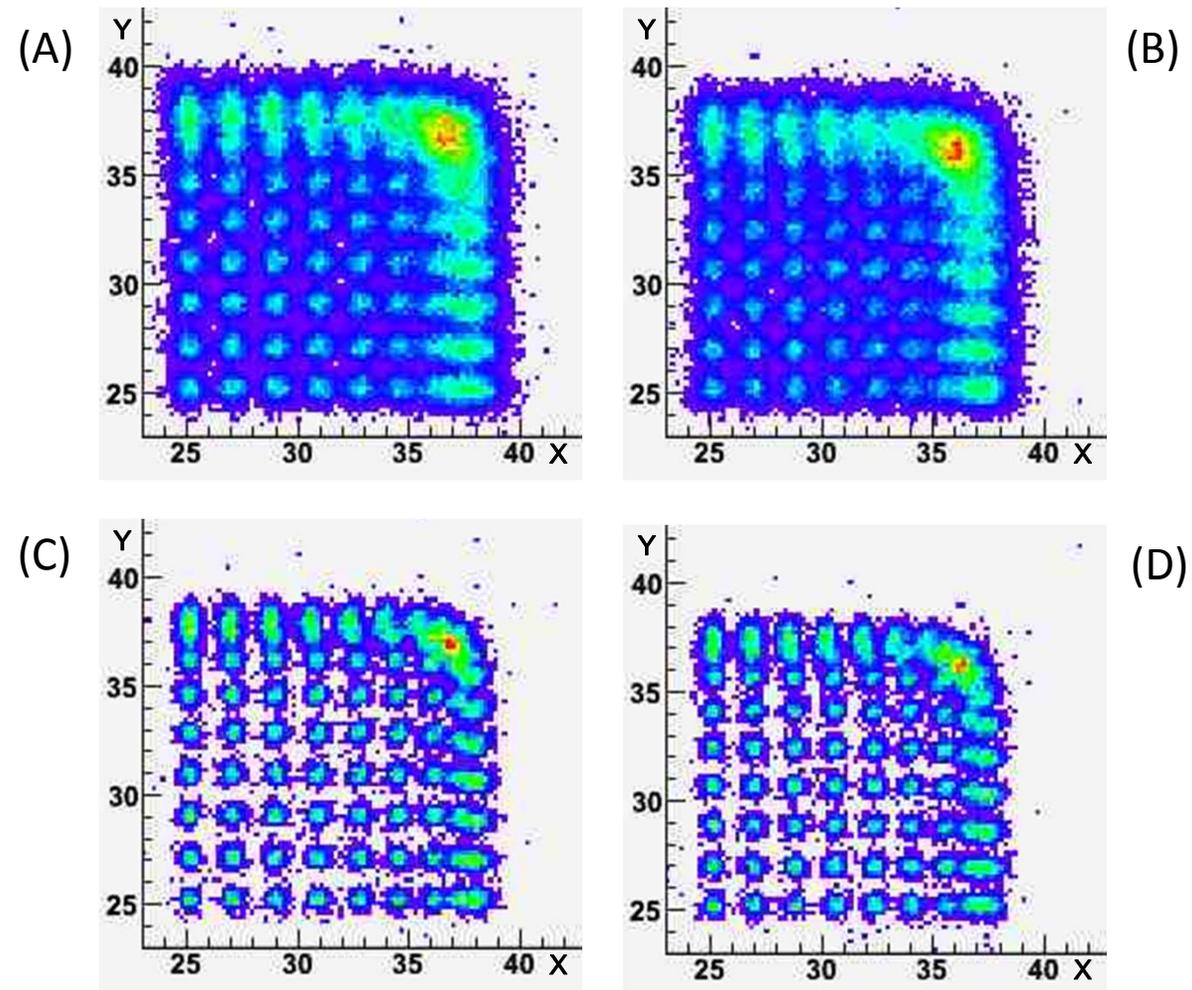

Figure 6

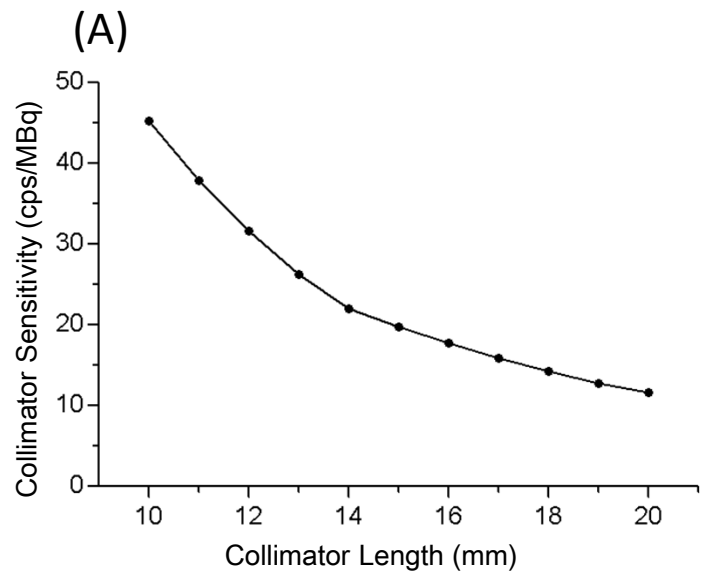

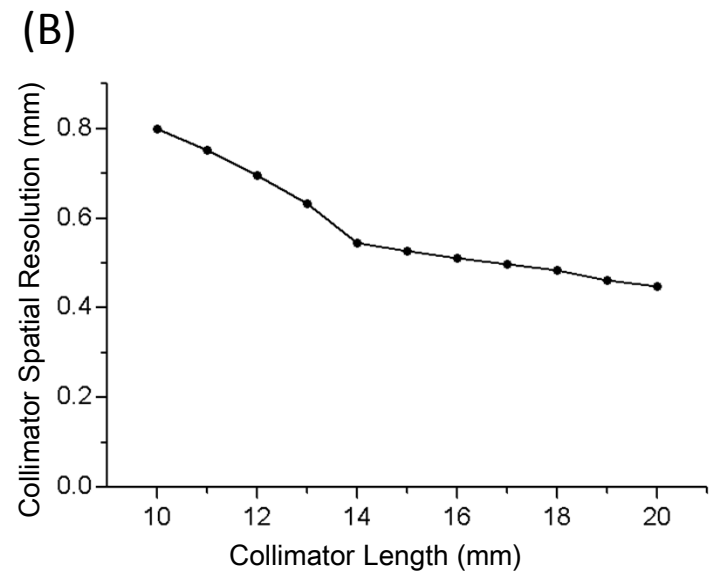

Figure 7

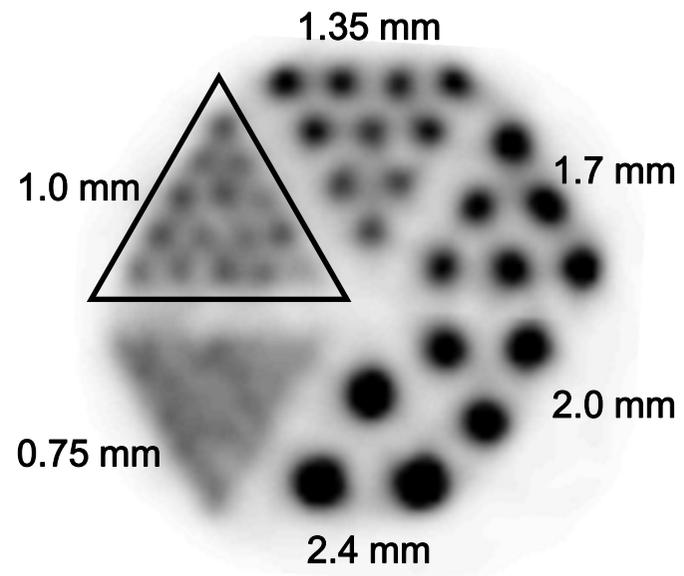